# A protected password change protocol

Ren-Chiun Wang[1], Chou-Chen Yang[2] and Kun-Ru Mo[3]

Department of Information Management[1]

Chaoyang University of Technology

168 Gifeng E.Rd., Wufeng, Taichung County, Taiwan 413, R.O.C.

Corresponding author

Email: chiunchiunwang@yahoo.com.tw

Department of Management Information System[2]

National Chung Hsing University

250 Kuo Kuang Road, 402 Taichung, Taiwan, R.O.C.

Email: cc_yang@nchu.edu.tw

Department of Information Management[3]

Min Chuan University

No. 250, section 5, Chung Shan North Road, Taipei City 111, Taiwan R.O.C.

Email: stephen@sanyo.com.tw

# A protected password change protocol

**Abstract**

Some protected password change protocols were proposed. However, the previous protocols were easily vulnerable to several attacks such as denial of service, password guessing, stolen-verifier and impersonation attacks etc. Recently, Chang et al. proposed a simple authenticated key agreement and protected password change protocol for enhancing the security and efficiency. In this paper, authors shall show that password guessing, denial of service and known-key attacks can work in their password change protocol. At the same time, authors shall propose a new password change protocol to withstand all the threats of security.

## 1. Introduction

Password is a short secret. Hence, it is easily be remembered by communication parties and can enhance the efficiency of the scheme. Many password-based authentication schemes were proposed [4-7]. However, the disadvantages of the scheme are easily vulnerable to the guessing, replay, denial of service and stolen-verifier attacks etc.

Beside that, a key agreement scheme should be considered into a password-based authentication scheme for enhancing the security. Recently, Peyravian and Zunic [8] proposed an authentication scheme for protecting password transmission by using one-way hash function [9]. In 2001, Tseng et al. [10] pointed out that their scheme was vulnerable to the server spoofing attack. At the same time, Tseng proposed an improvement based on the Diffie-Hellman scheme [2]. Not only that, many attacks and improvements were discussed continually [3-4, 11-13].

Recently, Chang et al. [1] proposed a simple authenticated key agreement and protected password change protocol for overcoming the above attacks. In this paper, authors shall show that Chang et al.'s password change protocol is vulnerable to the known-key, off-line password guessing and denial of service attacks. At the same time, authors shall propose a new password changing protocol to withstand all the threats of security..

Next section, authors shall review Chang et al.'s password change protocol and show that the protocol's weaknesses. In Section 3, authors shall propose a new password changing protocol. Finally, authors shall make a conclusion for this paper.

## 2. Review of Chang et al.'s password change protocol

In this section, authors shall go over Chang et al.'s password change protocol and show that their protocol's weaknesses as follows.

### 2.1 Chang et al.'s scheme

Let $p$ and $q$ are two large prime numbers. $g$ is a generator with order $q$ in GF($p$). Two communication parties, Alice and Bob, share a secret password *pw*.

Step 1: Alice selects a random number $a$ and computes $R_A = g^a$ mod $p$. Alice sends ($R_A \oplus pw \parallel R_A \oplus new\ pw$) to Bob.

Step 2: Bob sends ($R_B \parallel H(K_B, R_A)$) to Alice, where $b$ is chosen by Bob, $R_B = g^b$ mod $p$, $K_B = R_A^b$ mod $p$ and H() denotes a secure one-way hash function. Note that $R_A = (R_A \oplus pw) \oplus pw$ and $new\ pw = R_A \oplus (R_A \oplus new\ pw)$.

Step 3: Alice sends ($H(K_A, R_B) \oplus new\ pw$) to Bob, where $K_A = R_B^a$ mod $p$. And then Bob uses the recovered *new pw* to retrieve $H(K_A, R_B)$. If $H(K_A, R_B)$ is equal to $H(K_B, R_B)$, Bob accepts this new password *new pw*.

### 2.2 Weaknesses on Chang et al.'s protocol

***Off-line password guessing attack:*** The attacker can first intercept ($R_A \oplus pw \parallel R_A \oplus new\ pw$) from the communication channel in the Step 1. Then the attacker can guess two passwords *pw'* and *new pw'* and use the guessed values to get $R_A' = pw' \oplus (R_A \oplus pw)$ and $R_A'' = new\ pw' \oplus (R_A \oplus new\ pw)$. If $R_A' = R_A''$, the attacker gets the *pw* and *new pw* at the same time.

***Denial of service attack:*** The attacker can select a random number $c$ to compute ($R_A \oplus new\ pw \oplus c$) and send ($R_A \oplus pw \parallel R_A \oplus new\ pw \oplus c$) to Bob in the Step 1. Bob will get (*new pw* $\oplus c$) by computing $R_A \oplus (R_A \oplus new\ pw \oplus c)$ in the Step 2. Finally, Alice sends ($H(K_A, R_B) \oplus new\ pw$) to Bob in the Step 3. The attacker intercepts it and sends ($H(K_A, R_B) \oplus new\ pw \oplus c$) to Bob. Bob will accept (*new pw* $\oplus c$) is a new password because he uses (*new pw* $\oplus c$) to retrieve the true verifier $H(K_A, R_B)$.

***Known-key attack:*** Once the past session key $K_A$ or $K_B$ is compromised by the attacker, the attacker can get *new pw* and use it to impersonate Alice to communicate with Bob.
a. Once the past session key $K_A$ or $K_B$ is compromised, the attacker can compute $H(K_A, R_B) \oplus (H(K_A, R_B) \oplus new\ pw)$ to get *new pw* in the Step 3. Note that $K_A = K_B$.
b. Similarly, once the used password *pw* is compromised, the attacker can compute *pw*

($R_A \oplus pw$) $\oplus$ ($R_A \oplus$ *new pw*) to get new password *new pw* in the Step 1.

## 3. Our scheme

In this section, we shall a new password change protocol. The parameters are same as Chang et al.'s protocol.

### 3.1 A new password change protocol

Step 1: Alice $\to$ Bob: ($R_A \oplus pw \parallel R_B \oplus$ *new pw*)

Alice selects two random numbers $a$ and $b$ and computes $R_A = g^a$ mod $p$ and $R_B = g^b$ mod $p$. Alice sends ($R_A \oplus pw \parallel R_B \oplus$ *new pw*) to Bob.

Step 2: Bob $\to$ Alice: $R_C \parallel$ H($Key_1$, $R_A$, ($R_B \oplus$ *new pw*), $ID_B$, $ID_A$)

Bob sends ($R_C \parallel$ H($Key_1$, $R_A$, ($R_B \oplus$ *new pw*), $ID_B$, $ID_A$)) to Alice, where $c$ is chosen by Bob, $R_C = g^c$ mod $p$, $Key_1 = R_A^c = g^{ac}$ mod $p$, $ID_A$ and $ID_B$ are Alice and Bob's identities respectively. Note that $R_A = (R_A \oplus pw) \oplus pw$.

Step 3: Alice $\to$ Bob: H($Key_1$, $Key_2$, $R_C$, $ID_A$, $ID_B$)

Alice computes $Key_1 = R_C^a = g^{ca}$ mod $p$ and verifies whether H($Key_1$, $R_A$, ($R_B \oplus$ *new pw*), $ID_B$, $ID_A$) is correct or not. If it is true, Alice computes $Key_2 = R_C^b = g^{cb}$ mod $p$ and sends (H($Key_1$, $Key_2$, $R_C$, $ID_A$, $ID_B$)) to Bob.

Step 4: Bob can select a candidate *new pw'* to retrieve $R_B'$. And then Bob computes $Key_2' = R_B'^c = g^{b'c}$ mod $p$. If H($Key_1$, $Key_2'$, $R_C$, $ID_A$, $ID_B$) is equal to H($Key_1$, $Key_2$, $R_C$, $ID_A$, $ID_B$), Bob gets the correct new password *new pw*; otherwise, Bob can re-do this action until he gets the correct password. Note that the length of a new password is short, therefore this step has to be finished in a polynomial time. If it is not, Bob can reject it.

### 3.2 Security analysis

***Password guessing attack:*** The attacker can intercept the messages ($R_A \oplus pw \parallel R_B \oplus$ *new pw*), ($R_C \parallel$ H($Key_1$, $R_A$, ($R_B \oplus$ *new pw*), $ID_B$, $ID_A$)) and (H($Key_1$, $Key_2$, $R_C$, $ID_A$, $ID_B$)) from the communication channel. The attacker cannot make password guessing attack on our protocol even if the attacker guesses *pw* and *new pw* at the same time, because the attacker has no the related verification data and faces the discrete logarithm problem.

***Denial of service attack:*** The attacker uses a random number $c$ to replace with H($Key_1$, $Key_2$, $R_C$, $ID_A$, $ID_B$). Although Bob cannot reject it in no time, Bob still can find this weakness in a polynomial time.

***Known-key attack:*** If one of *Key₁* and *Key₂* is compromised, and the attacker intercepts the message (*R_A* ⊕ *pw* || *R_B* ⊕ *new pw*), (*R_C* || H(*Key₁*, *R_A*, (*R_B* ⊕ *new pw*), *ID_B*, *ID_A*)) and (H(*Key₁*, *Key₂*, *R_C*, *ID_A*, *ID_B*)), the attacker still cannot get the new password except all the session keys are compromised.

## 4. Conclusion

In this paper, authors show that Chang et al.'s protocol is vulnerable to the off-line password guessing, denial of service and known-key attacks respectively. Authors also propose a new password change protocol to withstand all the threats of security..